\documentstyle[12pt,epsfig, psfig]{article}
\begin{document}

\noindent Stockholm\\
USITP 03-03\\
April 2003\\

\vspace{1cm}

\begin{center}

{\Large GEOMETRY OF BLACK HOLE}

\vspace{5mm}

{\Large THERMODYNAMICS}

\vspace{1cm}

{\large Jan \AA man}\footnote{Email address: ja@physto.se.}

\

{\large Ingemar Bengtsson}\footnote{Email address: ingemar@physto.se. 
Supported by VR.}

\

{\large Narit Pidokrajt}\footnote{Email address: narit@physto.se.}

\

{\sl Stockholm University, AlbaNova\\
Fysikum\\
S-106 91 Stockholm, Sweden}

\vspace{8mm}

{\bf Abstract}

\end{center}

\vspace{5mm}

\noindent The Hessian of the entropy function can be thought of as a metric 
tensor on the state space. In the context of thermodynamical fluctuation 
theory Ruppeiner has argued that the Riemannian geometry of this metric 
gives insight into the underlying statistical mechanical system; the claim 
is supported by numerous examples. We study this geometry for some 
families of black holes. It is flat for the BTZ and Reissner--Nordstr\"om 
black holes, while curvature singularities occur for the 
Reissner--Nordstr\"om--anti--de Sitter and Kerr black holes. 

\newpage

{\bf 1. Introduction}

\vspace{5mm}

\noindent It has been argued that the Hessian of the thermodynamic entropy 
function S, that is the matrix  

\begin{equation} g_{ij} \equiv - \partial_i\partial_jS(X) \ , \end{equation}

\noindent can be thought of as a Riemannian metric on the state space in a 
meaningful way. We will refer to it as the Ruppeiner metric. It is clearly 
assumed that the coordinates $X^i$ form some 
preferred affine space; in ordinary thermodynamics they are chosen to be 
the extensive variables of the system. In this paper we will study this 
geometry for some important families of black holes, choosing the conserved 
charges $M$, $J$ and $Q$ as coordinates. The idea is that the Riemannian 
curvature in some sense measures the complexity of the underlying statistical 
mechanical model, which in this case is unknown but may well be in the process 
of being uncovered by progress in quantum gravity. The results are rather 
pleasing: We will find that the BTZ and Reissner--Nordstr\"om families have 
a flat thermodynamic geometry, while the Reissner--Nordstr\"om--anti-de 
Sitter and Kerr families exhibit curvature singularities. 

Our belief that this is a meaningful result rests on an analogy to 
thermodynamic fluctuation theory, where a similar claim was originally based 
simply on the observation that the thermodynamic geometry of the ideal 
gas is flat \cite{Ruppeiner1}. Let us recall the argument: Let $W$ be the 
number of (equiprobable) microstates consistent 
with a given macroscopical state. Boltzmann argued that the macroscopic 
entropy is given by 

\begin{equation} S = k\ln{W} \ . \end{equation}

\noindent Einstein rewrote this equation as 

\begin{equation} P \propto e^{\frac{1}{k}S} \ , \end{equation}

\noindent where $P$ is the probability that the given macrostate will be 
realized. We can Taylor expand the entropy around an equilibrium state, taking into 
account that the entropy has a maximum there, and introduce the Hessian matrix 

\begin{equation} g_{ij} \equiv - \partial_i\partial_jS(X) \ . \label{3} \end{equation}

\noindent Here $X$ stands for the $n$ extensive variables shifted so that they 
take the value zero at equilibrium. The matrix is positive definite if the 
entropy is concave. If we normalize the resulting probability distribution 
(and set $k = 1$) we obtain

\begin{equation} P(X) = \frac{\sqrt{g}}{(2{\pi})^{\frac{n}{2}}}
e^{- \frac{1}{2}g_{ij}X^iX^j} \end{equation}

\noindent as the probability distribution governing fluctuations around the 
equilibrium state. The pair correlation functions are then given by the 
contravariant metric tensor, 

\begin{equation} \langle X^iX^j\rangle = g^{ij} \ . \end{equation} 

\noindent In the derivation we assume that the fluctuations are small. 
So far everything is standard \cite{Landau}. It is important to realize 
that the physical situation here is a system described by the canonical 
(or grand canonical) 
ensemble, and moreover that one extensive parameter (typically the volume) 
has been set aside and used to give an appropriate physical dimension to 
$g_{ij}$. If this is not done the Gibbs--Duhem relation will imply that 
$g_{ij}$ has a null eigenvector. 

Ruppeiner \cite{Ruppeiner} argues that the Riemannian geometry of the metric 
tensor $g_{ij}$ carries information about the underlying statistical mechanical 
model of the system. In particular he argues that the metric is flat if and 
only if the statistical mechanical system is non--interacting, while curvature 
singularities are a signal of critical behaviour---more precisely of divergent 
correlation lengths. This viewpoint has received support from various 
directions \cite{Mrugala} \cite{Salamon} \cite{Brody}. Evidently the 
construction is related to the Fisher--Rao metric that is used 
in mathematical statistics, although it is fair to add that the Riemannian 
geometry of the Fisher--Rao metric does not play any significant role 
there---statistical geometry is rather more subtle \cite{Cencov}. 

The reason why Ruppeiner's arguments do not apply directly to black holes is 
that the thermodynamics of black holes exhibit some unfamiliar features which 
are in fact generic to systems with long range interactions in general, and 
to self--gravitating systems in particular \cite{Padmanabhan}. First 
we encounter negative specific heats, that is to say that the entropy is 
not a concave function. Second there are no extensive variables. Technically 
this means that the Ruppeiner metric will not be positive definite; on 
the other hand it will not have any null eigenvectors either. But it also 
means that the canonical ensemble does not exist, and that it is difficult 
to choose a physical dimension for the metric. Nevertheless we believe that 
the Ruppeiner geometry of black holes is telling us something; our justification 
is mainly the {\it a posteriori} one that once it has been worked out for 
some examples we will find an interesting pattern. For some 
further observations on the role of the Ruppeiner metric in 
black hole physics see Ferrara et al. \cite{Gibbons}. For background information 
on black hole thermodynamics see Davies \cite{Davies}. 

Some technical comments before we begin: Although the definition of the 
Ruppeiner metric depends on a preferred affine coordinate system we can 
afterwards transform ourselves to any 
coordinate system that we find convenient. Here we take note of a related 
construction due to Weinhold \cite{Weinhold}, who defined a metric in the 
energy representation through 

\begin{equation} g^{W}_{ij} \equiv \partial_i\partial_jM(S,N^a) \ . \end{equation}

\noindent We use $M$ to denote energy and $N^a$ to denote any other extensive 
variables. The entropy function is naturally a function of $M$ and $N^a$, so 
that in this notation the Ruppeiner metric is 

\begin{equation} g_{ij} = - \partial_i\partial_jS(M, N^a) \ . \end{equation}
 
\noindent Of course we can transform the Ruppeiner metric to the coordinate 
system used to define the Weinhold metric, and conversely. Interestingly the 
two metrics are conformally related \cite{Ihrig} \cite{Mrugala}: 

\begin{equation} ds^2 = g_{ij}dM^idM^j = \frac{1}{T}g^W_{ij}dS^idS^j \ , 
\label{7} \end{equation}

\noindent where $M^i = (M, N^a)$, $S^i = (S, N^a)$ and $T$ denotes the temperature 

\begin{equation} T = \frac{\partial M}{\partial S} \ . \end{equation}

\noindent Eq. (\ref{7}) often provides the most convenient way to compute 
the Ruppeiner metric. 

The organization of paper is as follows: In section 2 we consider the 
Reissner--Nordstr\"om and Reissner--Nordstr\"om--anti-de Sitter black holes 
in some detail. In section 3 we give a briefer treatment of the Kerr and 
BTZ black holes, and make some brief observations on the three dimensional 
Kerr--Newman family. Our conclusions are in section 4.

\vspace{1cm} 

{\bf 2. Reissner--Nordstr\"om black holes.}

\vspace{5mm}

\noindent We will describe one case in full detail, and we choose the 
Reissner--Nordstr\"om family of black holes for this purpose. They are 
spherically symmetric black holes carrying mass $M$ and charge $Q$. 
The event horizon is ruled by a Killing vector field whose norm is 

\begin{equation} ||{\xi}||^2 = - \frac{1}{r^2}\left( \frac{r^4}{l^2} 
+ r^2 - 2Mr + Q^2\right) \ , \label{15} \end{equation}

\noindent where $r$ is a natural radial coordinate chosen so that the area 
of a sphere at constant $r$ equals $4{\pi}r^2$. For later reference we 
have included a negative cosmological constant 

\begin{equation} {\lambda} = - \frac{3}{l^2} \end{equation}

\noindent but for the time being we set ${\lambda} = 0$, in which case 
the polynomial defining $||{\xi}||^2$ has two roots $r_+$ and $r_-$. These 
values of $r$ characterize the outer and inner event horizons, respectively. 
We find that 

\begin{equation} M = \frac{1}{2}(r_+ + r_-) \hspace{8mm} Q^2 = r_+r_- 
\hspace{8mm} S = r_+^2 \ , \end{equation}

\noindent where the entropy is one quarter of the area of the event horizon 
times Boltzmann's constant, 

\begin{equation} S = \frac{k}{4}A = k{\pi}r_+^2 = r_+^2 \ , \end{equation}

\noindent and we exercised our right to set $k = 1/{\pi}$. The extremal 
limit, beyond which the singularity becomes naked, occurs when the root $r_+$ 
is a double root. This happens at  

\begin{equation} Q^2 = M^2 \hspace{5mm} \Leftrightarrow \hspace{5mm} 
\frac{Q^2}{S} = 1 \ . \end{equation} 

\noindent The thermodynamics of these black holes is now defined by the 
fundamental relation 

\begin{equation} M = \frac{\sqrt{S}}{2}\left( 1 + \frac{Q^2}{S}\right) \ . 
\end{equation}

\noindent This is in the energy representation, which proves to be the 
most convenient one here---the Ruppeiner metric becomes quite unwieldy 
when expressed in terms of its natural coordinates. 

The Hawking temperature is 

\begin{equation} T = \frac{\partial M}{\partial S} = \frac{1}{4\sqrt{S}}\left( 
1 - \frac{Q^2}{S}\right) \end{equation}

\noindent and the electric potential is given by 

\begin{equation} {\Phi} = \frac{\partial M}{\partial Q} = \frac{Q}{\sqrt{S}} 
\ . \end{equation}

\noindent In its natural coordinates the Weinhold metric becomes 

\begin{equation} ds^2_W = \frac{1}{8S^{\frac{3}{2}}}\left( - \left(1 - 
\frac{3Q^2}{S}\right)dS^2 - 8QdQdS + 8SdQ^2\right) \ . \end{equation}

\noindent We observe that the component $g^W_{SS}$ vanishes and changes 
sign at 

\begin{equation} \frac{Q^2}{S} = \frac{1}{3} \ . \end{equation}

\noindent This implies that the specific heat $C_Q$ diverges and changes 
sign there. In ref. \cite{Davies} Davies argued that this implies that the 
system is undergoing a phase transition. We will see that this is not 
so---nothing special happens to the convexity of the energy function at 
this point. 

It is essential to use the coordinates $S, Q$ in the definition of the 
Weinhold metric. But once we have it it is convenient to introduce 
the new coordinate 

\begin{equation} u = \frac{Q}{\sqrt{S}} \ ; \hspace{8mm} - 1 \leq u 
\leq 1 \ . \end{equation}

\noindent The limits on the coordinate range are set by the fact that 
the black hole becomes extremal there. We now find that 

\begin{equation} ds^2_W = \frac{1}{8S^{\frac{3}{2}}}\left( - (1-u^2)dS^2 
+ 8S^2du^2\right) \ . \end{equation}

\noindent This is on diagonal form. 

In these coordinates the Ruppeiner metric is given by 

\begin{equation} ds^2 = \frac{1}{T}ds^2_W = - \frac{dS^2}{2S} + 4S\frac{du^2}{1-u^2} 
\ . \end{equation}

\noindent This metric is flat. To see this introduce new coordinates 

\begin{equation} {\tau} = \sqrt{2S} \hspace{8mm} \sin{\frac{\sigma}{\sqrt{2}}} 
= u \ . \end{equation}

\noindent The Ruppeiner metric now takes the form 

\begin{equation} ds^2 = - d{\tau}^2 + {\tau}^2d{\sigma}^2 \ , \end{equation}

\noindent which is recognizable as a timelike wedge in Minkowski space 
when described by Rindler coordinates. This seems to us to be a surprising result 
and provides some {\it a posteriori} justification for considering the 
Ruppeiner metric in the first place.  

\begin{figure}
\centering
\psfig{file=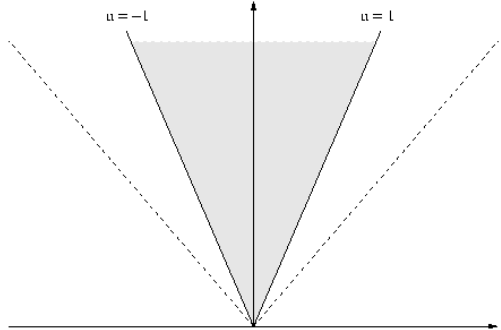,width=.5\textwidth}
\caption{The state space of the Reissner--Nordstr\"om black holes shown 
as a wedge in a flat Minkowski space.} 
\label{fig:RN-minkowski} 
\end{figure}

For completeness let us discuss the case of a non--zero and negative cosmological 
constant. Eq. (\ref{15}) is now a quartic polynomial. The event horizon is 
determined by its largest positive root $r_+$. The entropy is still determined 
by one quarter of the area of the event horizon, and it is not difficult to see 
that the fundamental relation is given by 

\begin{equation} M = \frac{\sqrt{S}}{2}\left( 1 + \frac{S}{l^2} + \frac{Q^2}{S}
\right) \ . \end{equation} 

\noindent The extremal limit occurs when $r_+$ is a degenerate root, and this 
happens when 

\begin{equation} \frac{Q^2}{S} = 1 + \frac{3S}{l^2} \ . \end{equation}

\noindent The Hawking temperature is 

\begin{equation} T = \frac{1}{4\sqrt{S}}\left( 1 + \frac{3S}{l^2} - \frac{Q^2}{S} 
\right) \ . \end{equation}

\noindent This vanishes in the extremal limit, as it should. 

The Weinhold metric is 

\begin{equation} ds^2_W = \frac{1}{8S^{\frac{3}{2}}}\left( - \left(1 - \frac{3S}{l^2} 
- \frac{3Q^2}{S}\right) dS^2 - 8QdSdQ + 8SdQ^2\right) \ . \end{equation}

\noindent It can be diagonalized using the same coordinate transformation as 
above, with the result that the conformally related Ruppeiner metric becomes 

\begin{equation} ds^2 = \frac{1}{1 + \frac{3{\tau}^2}{2l^2} - u^2}\left( 
- \left(1 - \frac{3{\tau}^2}{2l^2} - u^2\right) d{\tau}^2 + 2{\tau}^2du^2
\right) \ . \end{equation}

\noindent The geometry is non--trivial. By inspection we see that the signature 
of the metric---and with it the stability properties of the thermodynamic 
system---changes for sufficiently large black holes 
(using the length scale $l$ set by ${\lambda}$). This feature is of course 
well known---it means that the entropy function becomes concave for sufficiently 
large black holes \cite{Hawking}. The details of the thermodynamics of this case are 
actually quite interesting and can be found in the literature \cite{Louko} \cite{Myers}.
Our concern is the curvature scalar of the Ruppeiner metric, which is 

\begin{equation} R = \frac{9}{l^2}\frac{\left(\frac{3S}{l^2} + \frac{Q^2}{S}\right) 
\left( 1 - \frac{S}{l^2} - \frac{Q^2}{S}\right)}{\left( 1 - \frac{3S}{l^2} - 
\frac{Q^2}{S}\right)^2\left(1 + \frac{3S}{l^2} - \frac{Q^2}{S}\right) } 
\ . \end{equation}

\noindent We observe that the curvature diverges both in the extremal limit 
and along the curve where the metric changes signature, that is where the 
thermodynamical stability properties are changing. 

\vspace{1cm} 

\begin{center}
\begin{figure}
\begin{minipage}[t]{.32\textwidth}
   \includegraphics[width=\textwidth]{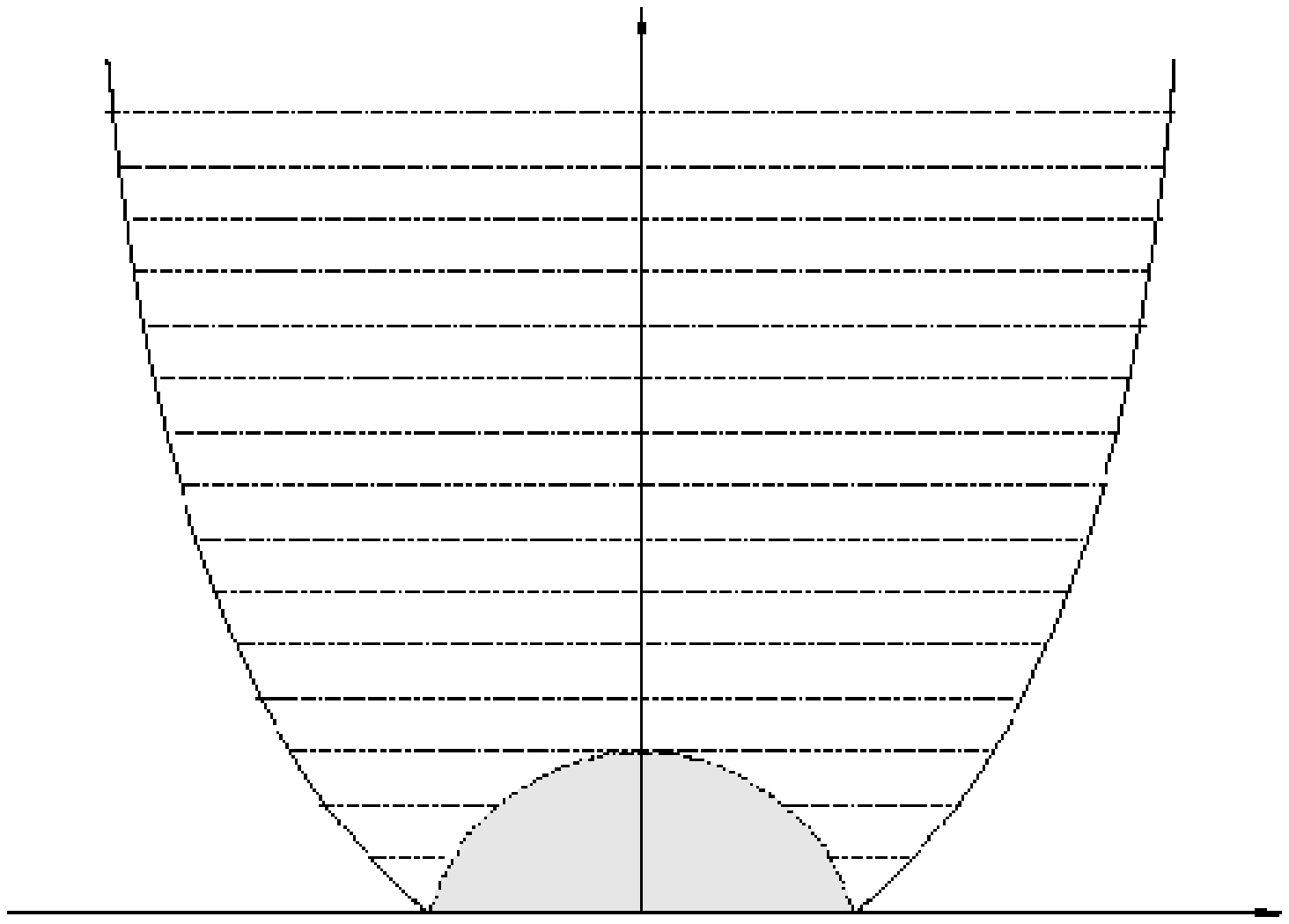}
\centering \texttt{Fig.(A)}
\end{minipage}
\begin{minipage}[t]{.32\textwidth}
   \includegraphics[width=\textwidth]{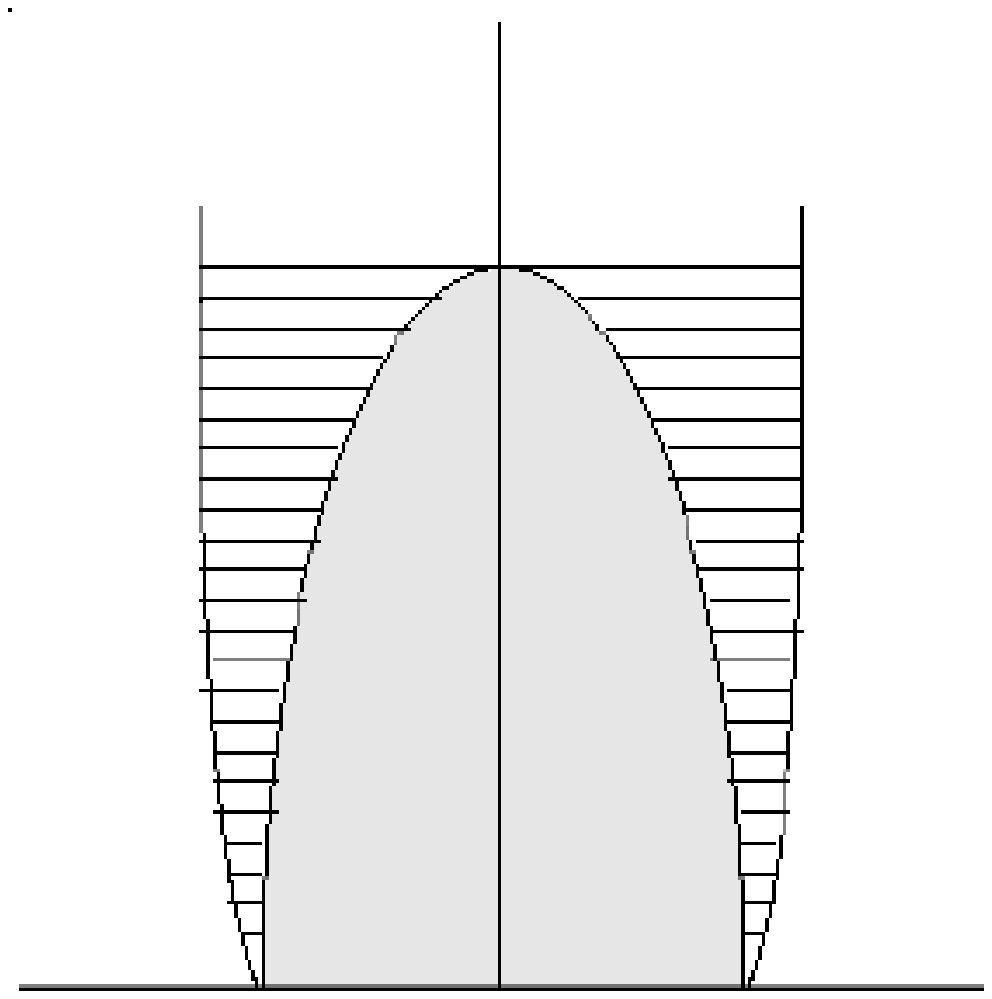}
\centering \texttt{Fig.(B)}
\end{minipage} 
\begin{minipage}[t]{.32\textwidth}
   \includegraphics[width=\textwidth]{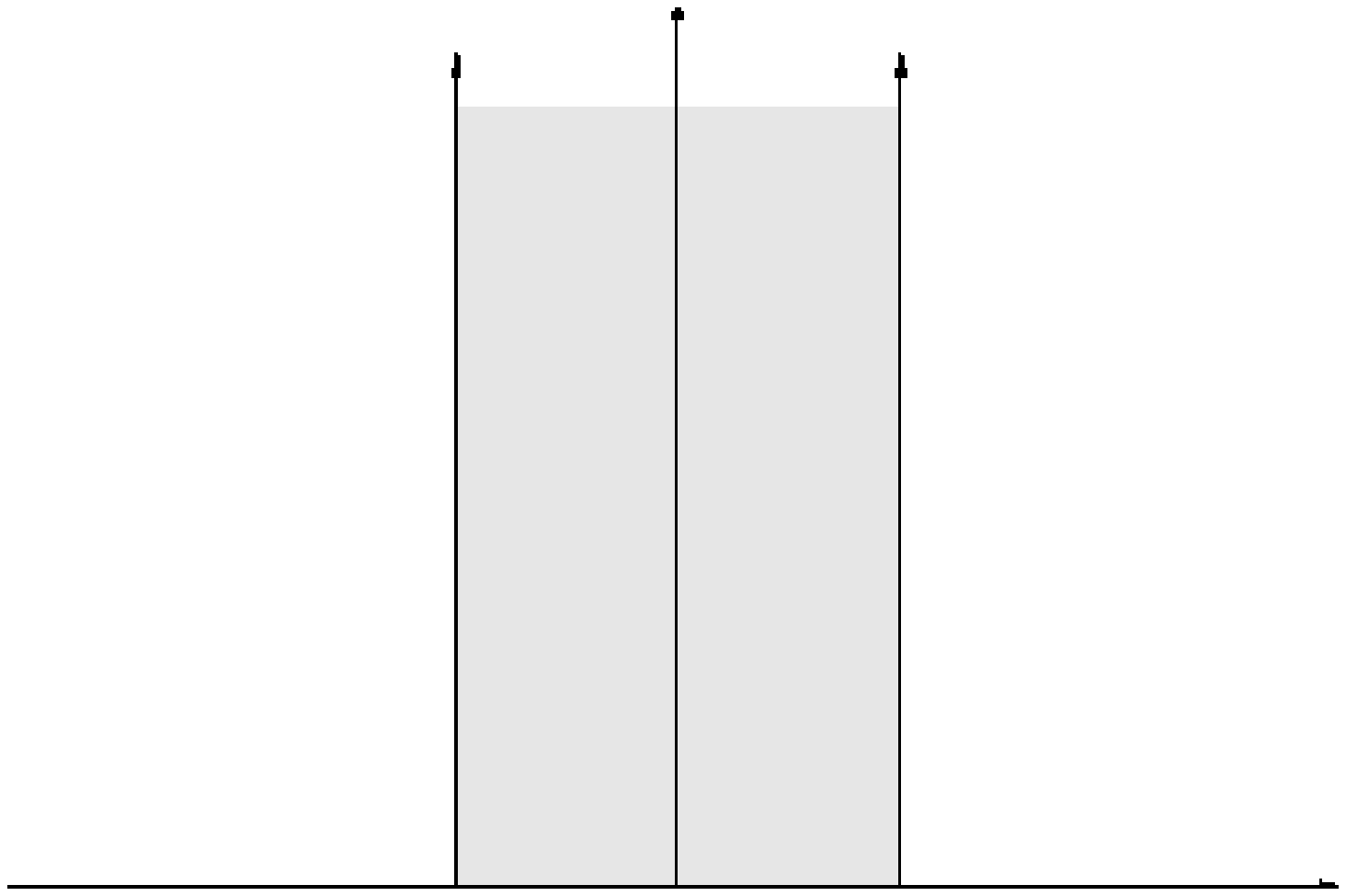}
\centering \texttt{Fig.(C)}
\end{minipage} 
\caption{The state space for Reissner--Nordstr\"om--anti-de Sitter black holes; 
our coordinates are $u$ and $S$ and the cosmological constant decreases as we 
go from A to C. The grey region has a Lorentzian metric. Note that C is the same 
as figure \ref{fig:RN-minkowski} although the coordinates differ.}
\label{fig:RNadS-figs} 
\end{figure}
\end{center}

{\bf 3. Other black holes.}

\vspace{5mm}

\noindent The Reissner--Nordstr\"om black holes belong to the three parameter 
Kerr--Newman family of black holes, with fundamental relation 

\begin{equation} M = \sqrt{\frac{S}{4} + \frac{1}{S}\left( J^2 + \frac{Q^4}{4}\right) 
+ \frac{Q^2}{2}} \ , \end{equation}

\noindent or in the entropy representation

\begin{equation} S = 2M^2 - Q^2 + 2M^2
\sqrt{1 - \frac{Q^2}{M^2} - \frac{J^2}{M^4}} \ . \end{equation}

\noindent Here $J$ measures 
the spin of the black hole. If we set $Q = 0$ we obtain the Kerr black 
holes, which are worthy of attention because they are believed to exist 
as physical objects. Their extremal limit is given by 

\begin{equation} \frac{J}{M^2} = \pm 1 \ . \end{equation}

\noindent From our point of view it is advantageous to use 
the entropy representation here. The Ruppeiner metric becomes 

\begin{equation} ds^2 = \frac{2}{(1 - \frac{J^2}{M^4})^{\frac{3}{2}}}
\left( - 2\left( (1 - \frac{J^2}{M^4})^{\frac{3}{2}} + 1 - \frac{3J^2}{M^4}
\right) dM^2 - \frac{4J}{M^3}dMdJ + \frac{dJ^2}{M^2}\right) . \end{equation}

\noindent This can be diagonalized by means of the coordinate transformation

\begin{equation} v = \frac{J}{M^2} \ ; \hspace{8mm} - 1 \leq v \leq 1 \ . 
\end{equation}

\noindent The Ruppeiner geometry is curved, but its curvature scalar takes 
a quite simple form: 

\begin{equation} R = \frac{1}{4M^2}\frac{\sqrt{1 - \frac{J^2}{M^4}} - 2}
{\sqrt{1 - \frac{J^2}{M^4}}} \ . \end{equation}

\noindent We observe that $R$ diverges in the extremal limit. It is however 
difficult to draw any firm conclusions from this because of the difficulty 
that the entropy function is not concave so that the fluctuation theory 
does not apply. A curious observation is that the Weinhold geometry 
of the Kerr black holes is actually flat. 

We used the computer program Classi \cite{Classi} to study the full three dimensional 
state space of the Kerr--Newman black holes (and to add some details to the table 
below). In particular, we computed the 
Cotton--York tensor and from this we could conclude that the Ruppeiner geometry 
is not conformally flat. Beyond this we did not uncover any noteworthy features.

There is one case where the thermodynamical response functions are positive 
throughout. This is the case 
of the 2+1 dimensional BTZ black holes \cite{BTZ}. They occur 
in a theory---Einstein's equation in 2+1 dimensions with a negative cosmological 
constant included---that is close to trivial from a dynamical point of view, 
but they are {\it bona fide} black holes nevertheless. Their thermodynamics 
is given by the fundamental relation

\begin{equation} M = S^2 + \frac{J^2}{4S^2} \ , \end{equation}

\noindent where we choose $k = 2/{\pi}$. The extremal limit, beyond which 
no black hole exists because the singularity (or ``singularity'', for 
connaisseaurs of these solutions \cite{BTZH}) becomes naked, is given by 
$J = \pm M$. It is also worth noting that $M = 0$ does not correspond to 
the ``background'' anti-de Sitter spacetime but to another kind of 
extremal black hole. 

This time the energy representation is the convenient one to use. The 
Weinhold metric diagonalizes if we trade $J$ for the new coordinate 

\begin{equation} u \equiv \frac{J}{2S^2} \ ; \hspace{1cm} - 1 \leq u \leq 1 \ . 
\end{equation}

\noindent Finally the Ruppeiner metric is 

\begin{equation} ds^2 = \frac{dS^2}{S} + \frac{Sdu^2}
{1 - u^2} \ . \end{equation}

\noindent This is a wedge of an Euclidean flat space described in coordinates 
that are polar coordinates in slight disguise. 

We summarize our results in a table.

\

\begin{tabular}{|l|l|l|} \hline\hline 
Black hole family & Ruppeiner & Weinhold \\ \hline
RN & Flat & Curved, no Killing vectors \\
RNadS & Curved, no Killing vectors & Curved, no Killing vectors \\
Kerr & Curved, no Killing vectors & Flat \\
BTZ & Flat & Curved, no Killing vectors \\ 
Kerr--Newman & Curved & Curved \\ \hline\hline
\end{tabular}

\vspace{1cm} 

{\bf 4. Conclusions}

\vspace{5mm}

\noindent In conclusion we have studied the Ruppeiner and Weinhold geometries 
of BTZ and Kerr--Newman black holes. In analogy to thermodynamic fluctuation 
theory we expect that a flat Ruppeiner geometry is a sign that an 
underlying statistical mechanical model must be exceptionally simple 
(``non--interacting''), while curvature singularities signal exceptional 
(``critical'') behaviour in the underlying model. We found that the 
Ruppeiner geometry is flat for the BTZ and Reissner--Nordstr\"om families, 
while the curvature diverges in the extremal limit in the Kerr and 
Reissner--Nordstr\"om--anti-de Sitter families. In the latter case the 
curvature is also singular along the line where the stability properties 
change. We find these results sensible, and also simpler than one would 
perhaps have expected. For the full Kerr--Newman family 
no elegant results were found. 

An interesting but of course very speculative 
use of the Ruppeiner geometry for the Kerr family is to let the volume form 
serve as a Bayesian prior for the amount of spin one should expect; observations 
concerning this can be expected in the not too far future.

\vspace{1cm} 

\end{document}